\documentclass[a4paper,11pt]{article}

\usepackage{enumerate}
\usepackage{amssymb,amsmath,epsfig,amsthm,color}
\usepackage{boxedminipage}

 \usepackage[ pdftex, plainpages = false, pdfpagelabels, 
                 bookmarks=false,
                 bookmarksopen = true,
                 bookmarksnumbered = true,
                 breaklinks = true,
                 linktocpage,
                 pagebackref,
                 colorlinks = true,  
                 linkcolor = blue,
                 urlcolor  = blue,
                 citecolor = red,
                 anchorcolor = green,
                 hyperindex = true,
                 hyperfigures
                 ]{hyperref}

\usepackage{vmargin}
\setmarginsrb{1.1in}{1.1in}{1.1in}{1.1in}{0mm}{0mm}{0mm}{7mm}

\newtheorem{rrule}{Reduction Rule}
\newtheorem{lemma}{Lemma}
\newtheorem{theorem}{Theorem}

\newcommand{\cc}{{\mathcal{C}}}
\newcommand{\Oh}{{\mathcal{O}}}
\newcommand{\lcac}{{\textbf{LCA-closure}}}
\newcommand{\hd}{\textbf{d}_{H}}
\newcommand{\cupall}{\pmb{\pmb{\cup}}}

\title{\bf Tree Deletion Set has a Polynomial Kernel
\newline
(but no OPT$^{\Oh(1)}$ approximation)
}

\author{Archontia C. Giannopoulou\thanks{Department of Informatics, University of Bergen, P.O. Box 7803, N-5020 Bergen, Norway.
\texttt{\{archontia.giannopoulou,daniello\}@ii.uib.no}.}~\thanks{This author's research, leading to these results, has received funding from the European Research Council under the European Union's Seventh Framework Programme (FP/2007-2013) / ERC Grant Agreement n. 267959.} 
\and Daniel Lokshtanov$^{*}$\thanks{Supported by Bergen Research Foundation grant BeHard.}
\and Saket Saurabh$^{*}$\thanks{The Institute of Mathematical Sciences, Chennai, India. \texttt{saket@imsc.res.in}. Supported by Parameterized Approximation, ERC Starting Grant 306992.}
\and Ond\v{r}ej Such\'{y}\thanks{Faculty of Information Technology, Czech Technical University Prague, Czech Republic. \texttt{ondrej.suchy@fit.cvut.cz}.}
}

\date{}
\begin{document}
\maketitle

\begin{abstract}
In the {\sc Tree Deletion Set} problem the input is a graph $G$ together with an integer $k$. The objective is to determine whether there exists a set $S$ of at most $k$ vertices such that $G \setminus S$ is a tree. The problem is {\sf NP}-complete and even {\sf NP}-hard to approximate within any factor of OPT$^c$ for any constant $c$. In this paper we give a $\Oh(k^4)$ size kernel for the {\sc Tree Deletion Set} problem. To the best of our knowledge our result is the first counterexample to the ``conventional wisdom'' that kernelization algorithms automatically provide approximation algorithms with approximation ratio close to the size of the kernel.
An appealing feature of our kernelization algorithm is a new algebraic reduction rule that we use to handle the instances on which {\sc Tree Deletion Set} is hard to approximate.
\end{abstract}

\section{Introduction} \label{sec_intro}
In the {\sc Tree Deletion Set} problem we are given as input an undirected graph $G$ and integer $k$, and the task is to determine whether there exists a set $S \subseteq V(G)$ of size at most $k$ such that $G \setminus S$ is a tree, that is a connected acyclic graph. This problem was first mentioned by Yannakakis~\cite{Yannakakis79} and is closely related to the classical {\sc Feedback Vertex Set} problem. Here input is a graph $G$ and integer $k$ and the goal is to decide whether there exists a set $S$ on at most $k$ vertices such that $G \setminus S$ is acyclic. The only difference between the two problems is that in {\sc Tree Deletion Set} $G \setminus S$ is required to be connected, while in {\sc Feedback Vertex Set} it is not. Both problems are known to be {\sf NP}-complete~\cite{GJ79,Yannakakis79}.

Despite the apparent similarity between the two problems their computational complexity differ quite dramatically. {\sc Feedback Vertex Set} admits a factor $2$-approximation algorithm, while {\sc Tree Deletion Set} is known to not admit any approximation algorithm with ratio $\Oh(n^{1-\epsilon})$ for any $\epsilon > 0$, unless {\sf P} = {\sf NP}~\cite{BafnaBF99,Yannakakis79}. With respect to parameterized algorithms, the two problems exhibit more similar behavior. Indeed, most techniques that yield fixed parameter tractable algorithms for {\sc Feedback Vertex Set}~\cite{ChenFLLV08,CaoCL10} can be adapted to also work for {\sc Tree Deletion Set}~\cite{RamanSS13}. 

It is also interesting to compare the behavior of the two problems with respect to polynomial time preprocessing procedures. Specifically, we consider the two problems in the realm of {\em kernelization}. We say that a parameterized graph problem admits a {\em kernel} of size $f(k)$ if there exists a polynomial time algorithm, called a {\em kernelization algorithm}, that given as input an instance $(G,k)$ to the problem outputs an equivalent instance $(G',k')$ with $k' \leq f(k)$ and $|V(G')| + |E(G')| \leq f(k)$. If the function $f$ is a polynomial, we say that the problem admits a {\em polynomial kernel}. We refer to the surveys~\cite{GN07,LokshtanovMS12} for an introduction to kernelization. For the {\sc Feedback Vertex Set} problem, Burrage et al.~\cite{BurrageEFLMR06} gave a kernel of size $\Oh(k^{11})$. Subsequently, Bodlaender~\cite{Bodlaender07} gave an improved kernel of size $\Oh(k^3)$  and finally Thomass\'{e}~\cite{Thomasse10} gave a kernel of size $\Oh(k^2)$. On the other hand the existence of a polynomial kernel for {\sc Tree Deletion Set} was open until this work. It seems difficult to directly adapt any of the known kernelization algorithms for {\sc Feedback Vertex Set} to {\sc Tree Deletion Set}. Indeed, Raman et al.~\cite{RamanSS13} conjectured that {\sc Tree Deletion Set} does not admit a polynomial kernel. 

The main reason to conjecture that {\sc Tree Deletion Set} does not admit a polynomial kernel stems from an apparent relation between kernelization and approximation algorithms. Prior to this work, all problems that were known to admit a polynomial kernel, also had approximation algorithms with approximation ratio polynomial in OPT. Here OPT is the value of the optimum solution to the input instance. In fact most kernelization algorithms are already approximation algorithms  with approximation ratio polynomial in OPT. This relation between approximation and kernelization led Niedermeier~\cite{rolf_invitation} to conjecture that {\sc Vertex Cover} does not admit a kernel with $(2-\epsilon)k$ vertices for $\epsilon > 0$, as this probably would yield a factor $(2-\epsilon)$ approximation for the problem thus violating the Unique Games Conjecture~\cite{KhotR08}. 

It is easy to show that an approximation algorithm for {\sc Tree Deletion Set} with ratio $\text{OPT}^{\Oh(1)}$ would yield an approximation algorithm for the problem with ratio $\Oh(n^{1-\epsilon})$ thereby proving {\sf P} = {\sf NP}. In particular, suppose {\sc Tree Deletion Set} had an $\text{OPT}^{c}$ algorithm for some constant $c$. Since the algorithm will never output a set of size more than $n$, the approximation ratio of the algorithm is upper bounded by $\min(\text{OPT}^c, \frac{n}{\text{OPT}}) \leq n^{1-\frac{1}{c+1}}$. This rules out approximation algorithms for  {\sc Tree Deletion Set} with ratio $\text{OPT}^{\Oh(1)}$, and makes it very tempting to conjecture that  {\sc Tree Deletion Set} does not admit a polynomial kernel. 

In this paper we show that  {\sc Tree Deletion Set} admits a kernel of size $\Oh(k^4)$. To the best of our knowledge this is the first example of a problem which does admit a polynomial kernel, but does not admit any approximation algorithm with ratio $\text{OPT}^{\Oh(1)}$ under plausible complexity assumptions.\\

\noindent {\bf Our methods.} 
The starting point of our kernel are known reduction rules for {\sc Feedback Vertex Set} adapted to our setting. By applying these graph theoretical reduction rules we are able to show that there is a polynomial time algorithm that given 
an instance $(G,k)$ of {\sc Tree Deletion Set} outputs an equivalent instance $(G',k')$ and a partition of $V(G')$ into sets $C_{m}$, $C_{g}$, and $I$ such that 
\begin{enumerate}\setlength\itemsep{-.7mm}
\item\label{introlem:clse1} $|C_{m}|= \Oh(k^{2})$, 
\item\label{introlem:clse2} $|C_{g}|= \Oh(k^{4})$, 
\item\label{introlem:clse3} $I$ is an independent set, and 
\item\label{introlem:clse4} for every $v\in I$, $N_{G'}(v)\subseteq C_{m}$ and $N_{G'}(v)$ is a double clique.
\end{enumerate}

\noindent Here a ``double clique'' means that for every pair $x$, $y$ of vertices in $N_{G'}(v)$, there are two edges between them. Thus we will allow $G'$ to be a multigraph, and consider a double edge between two vertices as a cycle. In order to obtain a polynomial kernel for {\sc Tree Deletion Set} it is sufficient to reduce the set $I$ to size polynomial in $k$.

For every vertex $v \in I$ and tree deletion set $S$ we know that $|N_{G'}(v) \setminus S| \leq 1$, since otherwise $G' \setminus S$ would contain a double edge. Further, if $v \notin S$ then $v$ has to be connected to the rest of 
$G' \setminus S$ and hence  $|N_{G'}(v) \setminus S| = 1$, implying that $v$ is a leaf in $G' \setminus S$. Therefore $G' \setminus (S \cup I)$ must be a tree. We can now reformulate the problem as follows.

For each vertex $u$ in $G' \setminus I$ we have a variable $x_u$ which is set to $0$ if $u \in S$ and $x_u = 1$ if $u \notin S$. For each vertex $v \in I$ we have a linear equation $\sum_{u \in N(v)} x_u = 1$. The task is to determine whether it is possible to set the variables to $0$ or $1$ such that (a) the subgraph of $G'$ induced by the vertices with variables set to $1$ is (connected) a  tree   and (b) the number of variables set to $0$ plus the number of unsatisfied linear equations is at most $k$.

At this point it looks difficult to reduce $I$ by graph theoretic means, as performing operations on these vertices correspond to making changes in a system of linear equations. In order to reduce $I$ we prove that there exists an algorithm that given a set ${\cal S}$ of linear equations on $n$ variables and an integer $k$ in time $\Oh(|{\cal S}| n^{\omega-1}k)$ outputs a set ${\cal S}'\subseteq {\cal S}$ of at most $(n+1)(k+1)$ linear equations such that any assignment of the variables that violates at most $k$ linear equations of ${\cal S}'$ satisfies all the linear equations of ${\cal S}\setminus {\cal S}'$. To reduce $I$ we simply apply this result and keep only the vertices of $I$ that correspond to linear equations in ${\cal S}'$. We believe that our reduction rule for linear equations adds to the toolbox of algebraic reduction rules for kernelization~\cite{KratschW12soda,KratschW12,Wahlstrom13} and will find more applications in the future.

\section{Basic Notions}

For every positive integer $n$ we denote by $[n]$ the set $\{1,2,\dots,n\}$ and for every set $S$ we denote by $\binom{S}{2}$ the 2-subsets of $S$. $\mathbb{N}$ denotes the natural numbers and $\mathbb{R}$ denotes the real numbers.

For a  {graph} $G=(V,E)$, we use  $V(G)$ to denote its vertex set $V$ and $E(G)$ to denote its edge set $E$.
If $S\subseteq V(G)$ we denote by $G\setminus S$ the graph obtained from $G$ after removing the vertices of $S$. 
In the case where $S=\{u\}$, we abuse notation and write $G\setminus u$ instead of $G\setminus \{u\}$.  
For $S\subseteq V(G)$, the {\em neighborhood} of $S$ in $G$, $N_{G}(S)$, is the set $\{u\in G\setminus S\mid \exists v\in S: \{u,v\}\in E(G)\}$.
Again, in the case where $S=\{v\}$ we abuse notation and write $N_{G}(v)$ instead of $N_{G}(\{v\})$.
We use  $\cc(G)$ to denote the set of the connected components of $G$.
Given a graph class ${\cal F}$ we denote by $\cupall {\cal F}$ the set $\cup_{G\in \mathcal{F}}V(G)$.
Given two vectors $x,y\in \mathbb{R}^{n}$, we denote by $\hd(x,y)$ the Hamming distance of $x$ and $y$, that is, $\hd(x,y)$ is equal to 
the number of positions where the vectors differ. For every $k\in \mathbb{N}$ we denote by $\mathbf{0}^{k}$ the vector $(0,0,\dots,0)\in \mathbb{R}^{k}$. 
When $k$ is implied from the context we abuse notation and denote $\mathbf{0}^{k}$ as $\mathbf{0}$.

Given a graph $G$ and a set $S\subseteq G$, we say that $S$ is a {\em feedback vertex set} of $G$ if the graph $G\setminus S$ does not contain any cycles. In the case
where $G\setminus S$ is connected we call $S$ {\em tree deletion set} of $G$.
Moreover, given a set $S\subseteq V(G)$, we say that $S$ is a double clique of $G$ if every pair of vertices in $S$ is joined by a double edge.

For a rooted tree $T$ and vertex set $M$ in $V(T)$ the least common ancestor-closure ({\em LCA-closure}) $\lcac(M)$ is obtained by the following process.
Initially, set $M'=M$. Then, as long as there are vertices $x$ and $y$ in $M'$ whose least common ancestor $w$ is not in $M'$, add $w$ to $M'$. Finally, output
$M'$ as the LCA-closure of $M$.

\begin{lemma}[\cite{FominLMS12}]\label{caclsbnd}
Let $T$ be a tree, $M\subseteq V(T)$, and $M'=\lcac(M)$. Then, $|M'|\leq 2|M|$ and for every connected component $C$ of $T\setminus M'$, $|N(C)|\leq 2$.
\end{lemma}

\section{A polynomial kernel for {\sc Tree Deletion Set}}
In this section we prove a polynomial size kernel for a weighted variant of the  {\sc Tree Deletion Set} problem. More precisely the problem we will study is following. 

\begin{center}
\begin{boxedminipage}{.99\textwidth}

\textsc{Weighted Tree Deletion Set  ({\sc wTDS})}

\begin{tabular}{ r l }
\textit{~~~~Instance:} & A graph $G$, a function $w:V(G)\rightarrow \mathbb{N}$, and\\
& a positive integer $k$. \\
\textit{Parameter:} & $k$.\\
\textit{Question:} & Does there exist a set $S\subseteq V(G)$ such that\\ 
& $\displaystyle \sum_{v\in S} w(v)\leq k$ and $G\setminus S$ is a tree? \\
\end{tabular}
\end{boxedminipage}
\end{center}

\subsection{Known Reduction Rules for {\sc wTDS}}

In this subsection we state some already known reduction rules for  {\sc wTDS} that are going to be needed during our proofs.

\begin{rrule}[\cite{RamanSS13}]\label{rrule1}
If $k<0$, then answer NO.
\end{rrule}

\begin{rrule}[\cite{RamanSS13}]
If the input graph is disconnected, then delete all vertices in connected components of weight less than $(\sum_{v\in V}w(v))-k$ 
and decrease $k$ by the weight of the deleted vertices.
\end{rrule}

\begin{rrule}[\cite{RamanSS13}]\label{rrule3}
If $v$ is of degree 1 and $u$ is its only neighbor, then delete $v$ and increase the weight of $u$ by the weight of $v$.
\end{rrule}

\begin{rrule}[\cite{RamanSS13}]\label{rrule4}
If $v_{0},v_{1},\dots, v_{l},v_{l+1}$ is a path in the input graph, such that $l\geq 3$ and $\deg(v_{i})=2$ for every $i\in [l]$, then 
$(a)$ replace the vertices $v_{1},\dots,v_{l}$ by two vertices $u_{1}$ and $u_{2}$ with edges $\{v_{0},u_{1}\}$, $\{u_{1},u_{2}\}$, 
and $\{u_{2},v_{l+1}\}$ and with $w(u_{1})=\min\{w(v_{i})\mid i\in [l]\}$ and $w(u_{2})=\left(\sum_{i=1}^{l}w(v_{i})\right)-w(u_{1})$.
Moreover, if $l\geq 2$ and $w(v_{0})>k$ or $w(v_{l+1})>k$, then apply $(a)$ and then $(b)$ delete $u_{2}$ and connect 
$u_{1}$ directly to $v_{l+1}$.
\end{rrule}

Given a vertex $x$ of $G$, an {\em $x$-flower of order $k$} is a set of cycles pairwise intersecting exactly in $x$.
If $G$ has a $x$-flower of order $k+1$, then $x$ should be in every tree deletion set of weight at most $k$ as otherwise we would need at least $k+1$
vertices to hit all cycles passing through $x$.
Thus the following reduction rule is safe.

\begin{rrule}\label{rrule5}
Let $(G,k)$ be an instance of {\sc wTDS}. If $G$ has a $x$-flower of order at least $k+1$, then remove $x$ and decrease the parameter $k$ by the weight of $x$. 
The resulting instance is $(G\setminus \{x\},k-w(x))$.
\end{rrule}

\begin{theorem}[\cite{Thomasse10}]\label{algosepvrt}
Let $G$ be a multigraph and $x$ be a vertex of $G$ without a self loop. Then in polynomial time we can either check whether there is 
a $x$-flower of order $k+1$ or find a set of vertices $Z\subseteq V(G)\setminus \{x\}$ of size at most $2k$ intersecting every cycle containing $x$.
\end{theorem}

\begin{rrule}\label{rrule6}
Let $(G,k)$ be an instance of {\sc wTDS}. If $v$ is a vertex such that $w(v)>k+1$, then let $w(v)=k+1$.
\end{rrule}

An instance $(G,k)$ of {\sc wTDS} is called {\em semi-reduced} if none of the Reduction Rules~\ref{rrule1} -~\ref{rrule6} can be applied.

\begin{theorem}[\cite{BafnaBF99}]\label{apprxalgofvs}
There is an $O(\min\{|E(G)|\log |V|,|V|^{2}\})$ time algorithm that given a graph $G$ that admits a feedback vertex set of size at most $k$ outputs a 
feedback vertex set of $G$ of size at most $2k$.
\end{theorem}

\subsection{A structural decomposition}\label{subsct:strcdecomp}

In this subsection we decompose an instance $(G,k)$ of {\sc wTDS} to an equivalent instance $(G',k)$ where $V(G')$ is partitioned into
three sets $C_{m}$, $C_{g}$, and $I$, such that the size of $C_{m}$ and $C_{g}$ is polynomial in $k$ and $I$ is an independent set.
In particular we obtain the following result. 

\begin{lemma}\label{mainlem}
There is a polynomial time algorithm that given a semi-reduced instance $(G,k)$ of {\sc wTDS} outputs an equivalent instance $(G',k')$ and 
a partition of $V(G')$ into sets $C_{m}$, $C_{g}$, and $I$ such that 
\begin{enumerate}[(i)]
\item\label{clse1} $|C_{m}|\leq 8k^{2}+2k$, 
\item\label{clse2} $C_{g}$ induces a forest and $|C_{g}|\leq 160k^{4}+248k^{3}+80k^{2}-16k-8$, 
\item\label{clse3} $I$ is an independent set, and 
\item\label{clse4} for every $v\in I$, $N_{G'}(v)\subseteq C_{m}$ and $N_{G'}(v)$ is a double clique.
\end{enumerate}
\end{lemma}

\begin{proof} We divide the proof into three parts: $(a)$ the identification of the set $C_{m}$ and the proof of~(\ref{clse1}), $(b)$ the transformation of $G$ to an 
intermediate graph $\widehat{G}$, the identification of the set $C_{g}$, and the proof of~(\ref{clse2}), 
and finally, $(c)$ the transformation of $\widehat{G}$ to $G'$, the identification of $I$ and the proof of~(\ref{clse3}) and~(\ref{clse4}).\\

\noindent {\bf Identification of the set $C_{m}$ and proof of~(\ref{clse1}).}
We begin the proof of the lemma with the identification of the set $C_{m}$. 
First notice that every tree deletion set of $G$ of weight at most $k$ is also a feedback vertex set of $G$ of size at most $k$ in the underlying non-weighted graph. Thus, by applying Theorem~\ref{apprxalgofvs} we may find 
in polynomial time a feedback vertex set $F$ of $G$. If $|F|> 2k$ then output NO. Otherwise,
 \begin{equation}
 |F|\leq 2k.\label{fvsszebnd}
 \end{equation} 
\noindent As the instance $(G,k)$ is semi-reduced, Reduction Rule~\ref{rrule5} is not applicable and $G$ does not contain a $x$-flower of order $k+1$, $x\in F$.
 Therefore, from Theorem~\ref{algosepvrt}, we get that for every $x\in F$ we can find in polynomial time a set $Q^{x}\subseteq V(G)\setminus \{x\}$ intersecting 
 every cycle that goes through $x$ in $G$ and such that 
\begin{equation} |Q^{x}|\leq 2k, x\in F.\label{sepszebnd}\end{equation}
Let $\displaystyle Q=\bigcup_{x\in F} Q^{x}$ and notice that from the definition of $Q$, and Eq.~\eqref{fvsszebnd} and~\eqref{sepszebnd}, 
\begin{equation}
\displaystyle \left|Q\right|\leq \left|\bigcup_{x\in F} Q^{x}\right| \leq\sum_{x\in F}\left|Q^{x}\right|\leq \sum_{x\in F}2k  \leq 4k^{2}.\label{qszebnd}
\end{equation}
 Let $\cc(G\setminus F)=\{H_{1},H_{2},\dots, H_{l}\}$ and note that, as $F$ is a feedback vertex set of $G$, all $H_{i}$'s are trees. 
 From now on, without loss of generality we will assume that $H_{i}$ is rooted at some vertex $v_{i}\in V(H_{i})$, $i\in [l]$.\\

\noindent Let $Q_{i}=V(H_{i})\cap Q$, $i\in [l]$. In other words, $Q_{i}$ denotes the set of vertices of $H_{i}$ that are also vertices of $Q$, $i\in [l]$. 
Let also $\widehat{Q}_{i}=\lcac(Q_{i})$, that is, let $\widehat{Q}_{i}$ denote the least common ancestor-closure of the set $Q_{i}$ in the tree $H_{i}$. 
Let $\displaystyle\widehat{Q}=\bigcup_{i\in [l]}\widehat{Q}_{i}$ and note that $\widehat{Q}\cap F=\emptyset$. Observe now that from 
Lemma~\ref{caclsbnd} and Eq.~\eqref{qszebnd} we get that
\begin{equation}\label{qclsszebnd}
\displaystyle \left|\widehat{Q}\right|=\left|\bigcup_{i\in [l]}\widehat{Q}_{i}\right|\leq \sum_{i\in [l]}\left|\widehat{Q}_{i}\right|\leq 
2\sum_{i\in l}\left|Q_{i}\right|\leq 2\left|Q\right|\leq 8k^{2}.
\end{equation}
Finally, we define $C_{m}$ to be the set $\widehat{Q}\cup F$.
From Eq.~\eqref{fvsszebnd} and~\eqref{qclsszebnd} it follows that 
\begin{equation}
|C_{m}|\leq 8k^{2}+2k.
\end{equation}
and we conclude the first part of our proof.\\

\noindent{\bf Transformation of $G$ to an intermediate graph $\widehat{G}$, identification of the set $C_{g}$, and proof of~(\ref{clse2}).}
We continue our proof by working towards the identification of the set $C_{g}$. 
First notice that a straightforward implication of Lemma~\ref{caclsbnd} is that we may partition $\cc(G\setminus C_{m})$
into three sets $\cc_{0}(G\setminus C_{m})$, $\cc_{1}(G\setminus C_{m})$, and $\cc_{2}(G\setminus C_{m})$ in such a way that $\cc_{i}(G\setminus C_{m})$ 
contains all the graphs in $\cc(G\setminus C_{m})$ that have exactly $i$ neighbors in $\widehat{Q}$, $0\leq i \leq 2$.\\

\noindent {\em Claim 1.} For every connected component $H\in \cc(G\setminus C_{m})$ and every vertex $y\in C_{m}$, $|N_{G}(y)\cap V(H)|\leq 1$, that is, 
every vertex $y$ of $F$ and every vertex $y$ of $\widehat{Q}$ have at most one neighbor in every connected component $H$ of $G\setminus C_{m}$.\\

\noindent {\em Proof of Claim 1.} Let $y\in C_{m}$ and $H\in \cc(G\setminus C_{m})$ and assume to the contrary that $|N_{G}(y)\cap V(H)|\geq 2$. 
Then, as $H$ is connected, the graph $G[V(H)\cup\{y\}]$ contains a cycle that goes through $y$.
 If $y\in F$, we end up to a contradiction to the facts that $G[V(H)\cup \{y\}]\subseteq G\setminus Q^{y}$
 and the set $Q^{y}$ intersects every cycle that goes through $y$. If $y\in \widehat{Q}$, it contradicts to the facts that 
 $G[V(H)\cup\{y\}]\subseteq G\setminus F$ (recall that $\widehat{Q}\cap F=\emptyset$) and $G\setminus F$ is acyclic. \hfill $\diamond$\\

\noindent Let ${\cal P}=\binom{C_{m}}{2}\setminus \binom{\widehat{Q}}{2}$, that is, 
${\cal P}=\{\{y,y'\}\subseteq C_{m} \mid |\{y,y'\}\cap \widehat{Q}|\leq 1\}$. For every pair $\{y,y'\}\in {\cal P}$ we let 
\begin{equation}
S(y,y') = \{H\in \cc_{0}(G\setminus C_{m})\cup \cc_{1}(G\setminus C_{m})\mid \{y,y'\}\subseteq N_{G}(V(H)) \}
\end{equation}
and
\begin{equation}
c(y,y') = |S(y,y')|.
\end{equation}
In other words, $S(y,y')$ is the set of the connected components in $\cc_{0}(G\setminus C_{m})\cup \cc_{1}(G\setminus C_{m})$ 
whose neighborhood in $G$ contains both $y$ and $y'$ and $c(y,y')$ denotes the cardinality of $S(y,y')$. We partition ${\cal P}$ into 
two sets ${\cal P}^{\leq k+1}$ and ${\cal P}^{\geq k+2}$. In particular
\begin{eqnarray}
{\cal P}^{\leq k+1} = \{\{y,y'\}\in {\cal P}\mid c(y,y')\leq k+1\}\\
{\cal P}^{\geq k+2} = \{\{y,y'\}\in {\cal P}\mid c(y,y')\geq k+2\}
\end{eqnarray}

\noindent Notice that Claim 1 implies that $c(y,y')$ equals the number of vertex-disjoint 
paths between $y$ and $y'$ whose internal vertices belong to  $\cc_{0}(G\setminus C_{m})\cup \cc_{1}(G\setminus C_{m})$.
Let $\{y,y'\}\in {\cal P}^{\geq k+2}$. 
Observe then that if neither $y$ nor $y'$ belong to a tree deletion set of $G$ we need at least $k+1$ vertices to hit all the cycles 
of $\cupall S(y,y') \bigcup \{y,y'\}$ as otherwise, from the Pigeonhole Principle, for every $S\subseteq \cupall S(y,y')$ with $|S|\leq k$ 
there exist two connected components $H_{1}$ and $H_{2}$ in $S(y,y')$ 
such that $(V(H_{1})\cup V(H_{2}))\cap S=\emptyset$ and thus the graph induced by $V(H_{1})\cup V(H_{2})\cup \{y,y'\}$ contains a cycle.
This implies that $(G,k)$ is a yes instance if and only if at least one of the vertices $y$ and $y'$ is contained
in every tree deletion set of $G$ of weight $k$. Thus, the instance $(\widehat{G},k)$ obtained from $G$ after adding
double edges between the pairs of vertices $y$ and $y'$, $\{y,y'\}\in {\cal P}^{\geq k+2}$, is equivalent to $(G,k)$ and $V(\widehat{G})=V(G)$.
This completes the transformation of $G$ to the intermediate graph $\widehat{G}$. We continue now with the identification of the set $C_{g}$
and the proof of~(\ref{clse2}).\\

\noindent We let ${\cal S}=\bigcup\{S(\{y,y'\})\mid \{y,y'\}\in {\cal P}^{\leq k+1}\}$ and finally, we define $C_{g}$ to be the set 
\begin{equation}\label{grbgdfn}
C_{g} = \cupall {\cal S}\bigcup  \cupall \cc_{2}(G\setminus C_{m}).
\end{equation}
\noindent From the definition, it follows that $C_{g}$ induces a forest. We now prove the upper bound on the size of $C_{g}$. We first need the following.\\

\noindent {\em Claim 2.}
If $H$ is a connected component of $\displaystyle G\setminus C_{m}$ then $|V(H)|\leq 8k+8$.\\

\noindent {\it Proof of Claim 2.}
Let $H$ be a connected component of $\displaystyle G\setminus C_{m}$. 
First recall that, from Claim 1, every vertex of $C_{m}$ has at most 1 neighbor in $H$ and, from construction of $\widehat{Q}$, 
$H$ has at most 2 neighbors in $\widehat{Q}$. This implies that there are at most $|F|+2\leq 2k+2$ vertices in $H$ that have a 
neighbor in $G\setminus V(H)$, and in particular in $C_{m}$. We call this set of vertices $N$.
Let $H_{1}$ be the set of vertices of degree 1 in $H$, that is, the leaves of $H$. From Reduction Rule~\ref{rrule3} it follows that for every $v\in H_{1}$, $\deg_{G}(v)\geq 2$ and thus, as $H\in \cc(G\setminus C_{m})$, $v$ has at least one neighbor in $C_{m}$. Therefore, $H_{1}\subseteq N$ and 
$$|H_{1}|\leq 2k+2.$$
Let now $H_{3}$ be the set of vertices of degree at least 3 in $H$. For $H_{3}$ it is easy to observe that, by standard combinatorial arguments on trees, 
$$|H_{3}|\leq |H_{1}|\leq 2k+2.$$

Finally, let $P$ be the set $N\cup H_{3}$ and ${\cal E}$ be the set of paths in $H$ with endpoints in $P$. Again, as $|P|\leq 4k+4$, it holds
that $|{\cal E}|\leq 2k+2$.
Observe that by construction of ${\cal E}$ all the inner vertices of the paths in ${\cal E}$ have degree exactly 2. Therefore, from Reduction Rule~\ref{rrule4} we get that every path in ${\cal E}$ contains at most 2 vertices. This implies that $|V(H)\setminus P|\leq 4k+4$.
To conclude, as $|V(H)|=|V(H)\setminus P|+|P|$, we get that $|V(H)|\leq 8k+8$.
 \hfill $\diamond$\\

\noindent We now prove an upper bound on $|\cupall {\cal S}|$. 
Notice first that, by definition, ${\cal S}$ can be written as the union of the graph classes ${\cal S}_{1}$ and ${\cal S}_{2}$, where 
\begin{eqnarray*}
{\cal S}_{1} & = & \bigcup_{\{x,y\}\in F \cap {\cal P}^{\leq k+1}}\left(H\in\cc_{0}(G\setminus C_{m})\cup\cc_{1}(G\setminus C_{m})\mid 
\{x,y\}\subseteq N_{G}(H)\right)
\end{eqnarray*}
and
\begin{eqnarray*}
{\cal S}_{2} & = & \bigcup_{x\in F}\left(\bigcup_{\substack{y\in \widehat{Q}\\ \{x,y\}\in {\cal P}^{\leq k+1}}} 
 \left(H\in\cc_{0}(G\setminus C_{m})\cup\cc_{1}(G\setminus C_{m})\mid \{x,y\}\subseteq N_{G}(H)\right)\right).
\end{eqnarray*}
Moreover, $\cupall {\cal S} =\cupall {\cal S}_{1} \cup \cupall {\cal S}_{2}$ and thus, it follows that 
\begin{equation}\label{eqngergkmlerg}
|\cupall {\cal S}|\leq |\cupall {\cal S}_{1}| + |\cupall {\cal S}_{2}|\leq |{\cal S}_{1}|\max\{|V(H)|\mid H\in {\cal S}_{1}\}+|{\cal S}_{2}|\max\{|V(H)|\mid H\in {\cal S}_{2}\}.
\end{equation}

\noindent From Claim 2, in order to prove an upper bound on $|\cupall {\cal S}|$ it is enough to prove upper bounds on $|{\cal S}_{1}|$ and $|{\cal S}_{2}|$.
Recall first that for every pair $\{x,y\}\in F\cap {\cal P}^{\leq k+1}$ there exist at most $k+1$ connected components containing both $x$ and $y$ in their
common neighborhood and therefore 
\begin{equation}\label{aefalwefnlkflwa}
\displaystyle |{\cal S}_{1}|\leq \binom{|F|}{2}(k+1)\leq \binom{2k}{2}(k+1)=2k^{3}+k^{2}-k.
\end{equation}
\noindent For the upper bound on ${\cal S}_{2}$, for every $x\in F$ we partition the set $\widehat{Q}$
 into two sets $R_{x}^{\leq 1}$ and $R_{x}^{\geq 2}$ in the following way.
\begin{eqnarray*}
R_{x}^{\leq 1} & = & \{y\in \widehat{Q}\mid \{x,y\}\in {\cal P}^{\leq k+1} \text{ and there exists at most one graph}\nonumber \\
& &  H\in \cc_{1}(G\setminus C_{m}) \text{ such that } \{x,y\}\subseteq N_{G}(H)\}\\
R_{x}^{\geq 2} & = & \{y\in \widehat{Q}\mid \{x,y\}\in {\cal P}^{\leq k+1} \text{ and there exist at least two distinct graphs}\nonumber \\
& & H_{1},H_{2}\in \cc_{1}(G\setminus C_{m}) \text{ such that } \{x,y\}\subseteq N_{G}(H_{1})\cap N_{G}(H_{2})\}.
\end{eqnarray*}
Observe that 
\begin{equation}\label{eqnealwefaw}
|{\cal S}_{2}|\leq \sum_{x\in F} \left(|R_{x}^{\leq 1}| + |R_{x}^{\geq 2}|(k+1)\right).
\end{equation} 
For every $x\in F$, it trivially holds that 
\begin{equation}\label{regaliuhgawel}
|R_{x}^{\leq 1}|\leq |\widehat{Q}|\leq 8k^{2}.
\end{equation}
Moreover, we claim that for every $x\in F$, $|R_{x}^{\geq 2}|\leq k$. Indeed, assume to the contrary that $|R_{x}^{\geq 2}|\geq k+1$ for some $x\in $F.
Then there exist $k+1$ vertices $y_{i}\in \widehat{Q}$, $i\in [k+1]$, such that
for every $i$ there exist two connected components $H_{1}^{i}$ and $H_{2}^{i}$ in $\cc_{1}(G\setminus C_{m})$ such that 
$\{x,y\}\subseteq N_{G}(H_{1}^{i})\cap N_{G}(H_{2}^{i})$. This implies that the graph induced by the vertex $x$, the vertices $y_{i}$, $i\in [k+1]$, and the
graphs $H_{1}^{i}$ and $H_{2}^{i}$, $i\in [k+1]$, contains a $x$-flower of order $k+1$ (notice that, as all the graphs belong to $\cc_{1}(G\setminus C_{m})$, 
they are pairwise disjoint). This is a contradiction to the fact that $G$ is semi-reduced.
Therefore, for every $x\in F$,
\begin{equation}\label{lwekfjwaegfa}
|R_{x}^{\geq 2}|\leq k.
\end{equation}
From Eq.~\eqref{eqnealwefaw},~\eqref{regaliuhgawel}, and~\eqref{lwekfjwaegfa} we obtain that
\begin{equation}\label{aergwalag}
|{\cal S}_{2}|\leq 18k^{3}+2k^{2}.
\end{equation}
Finally, from Eq.~\eqref{eqngergkmlerg},~\eqref{aefalwefnlkflwa},~\eqref{aergwalag}, and Claim 2 we get that.
\begin{equation}\label{rstsze}
|\cupall {\cal S}| \leq 160k^{4} + 184k^{3} + 16k^{2} - 8k.
\end{equation}

\noindent We continue by showing that 
\begin{equation}\label{cc2szebnd}
\left|\cupall \cc_{2}(G\setminus C_{m})\}\right|\leq 64k^{3}+64k^{2}-8k-8.
\end{equation}
\noindent Observe that, from Claim 2, in order to find an upper bound on $\left|\cupall \cc_{2}(G\setminus C_{m})\}\right|$ it is enough to 
find an an upper bound on $|\cc_{2}(G\setminus C_{m})|$.
We do so with the following claim.\\

\noindent {\em Claim 3.} $|\cc_{2}(G\setminus C_{m})|\leq |\widehat{Q}|-1\leq 8k^{2}-1$.\\

\noindent {\em Proof of Claim 3.} Let $\displaystyle C=\bigcup_{H\in \cc_{2}(G\setminus C_{m})} (N_{G}(H)\cap \widehat{Q})$, 
that is, let $C$ be the set of neighbors of the graphs in $\cc_{2}(G\setminus C_{m})$ in $\widehat{Q}$.
Let $A_{C}$ be the graph with vertex set $C$ where two vertices in $C$ are connected by an edge if and only if they are the neighbors of
a graph $H\in \cc_{2}(G\setminus C_{m})$ in $\widehat{Q}$. Hence, the number of edges of $A_{C}$ equals $|\cc_{2}(G\setminus C_{m})|$.
Notice then that if $A_{C}$ is a forest the claim follows from Eq.~\eqref{qclsszebnd}.
We now work towards showing that $A_{C}$ is a forest. Indeed, assume to the contrary that there exists a cycle in $A_{C}$. 
Then it is easy to see that we may find a cycle
in the graph $\widehat{H}$ induced by the graphs in $\cc_{2}(G\setminus C_{m})$ which correspond to the edges of the cycle in $A_{C}$
and their neighborhood in $\widehat{Q}$.
Recall that $\widehat{Q}\cap F=\emptyset$ and therefore $\widehat{H}$ is a subgraph of $G\setminus F$.
This contradicts to the fact that $F$ is a feedback vertex set of $G$ and completes the proof of the claim.
\hfill $\diamond$\\

\noindent To conclude the proof of the upper bound on $|C_{g}|$ we notice that Eq.~\eqref{cc2szebnd} follows from Claims 2 and 3, and the inequality below.
\begin{eqnarray*}
\bigcup |\{V(H)\mid H\in \cc_{2}(G\setminus C_{m})\}| & \leq & |\cc_{2}(G\setminus C_{m})|\cdot \max_{H\in \cc_{2}(G\setminus C_{m})}|V(H)|.
\end{eqnarray*}
Thus, from Eq.~\eqref{grbgdfn},~\eqref{rstsze}, and~\eqref{cc2szebnd}, we obtain that $|C_{g}|\leq 160k^{4}+248k^{3}+80k^{2}-16k-8$.

\medskip

\noindent {\bf Transformation of $\widehat{G}$ to $G'$, identification of $I$, and proof of~(\ref{clse3}) and~(\ref{clse4}).}
We now continue with the construction of the graph $G'$ and the identification of the set $I$.\\

\noindent In order to construct the graph $G'$ we first need the following claim for the connected components of $\widehat{G}\setminus (C_{m}\cup C_{g})$.\\

\noindent {\em Claim 4.} For every connected component $H\in \cc(\widehat{G}\setminus (C_{m}\cup C_{g}))$, $N_{\widehat{G}}(V(H))\subseteq C_{m}$ and
$N_{\widehat{G}}(V(H))$ is a double clique.\\

\noindent {\em Proof of Claim 4.} Recall that, $\cupall \cc_{2}(G\setminus C_{m})\subseteq C_{g}$ and that $C_{g}\subseteq \cupall\cc(G\setminus C_{m})$. It follows that
$\cc(\widehat{G}\setminus (C_{m}\cup C_{g}))\subseteq \cc(G\setminus C_{m})\setminus \cc_{2}(G\setminus C_{m})$. 
This implies that for every connected component $H\in \widehat{G}\setminus (C_{m}\cup C_{g})$, $N_{\widehat{G}}(H)\subseteq C_{m}$.
Let now $y$ and $y'$ be two vertices in ${\cal P}$ that are not joined by a double edge. 
By construction of the graph $\widehat{G}$, this implies that $\{y,y'\}\in {\cal P}^{\leq k+1}$.
Moreover, the definitions of the sets ${\cal S}$ and $C_{g}$, imply that there is no graph $H\in \cc(\widehat{G}\setminus (C_{m}\cup C_{g}))$ such that 
$\{y,y'\}\subseteq N_{C_{m}}(V(H))$. Therefore, for every graph $H\in \cc(\widehat{G}\setminus (C_{m}\cup C_{g}))$, 
the graph induced by $N_{\widehat{G}}(V(H))$ is a double clique.
 \hfill $\diamond$\\

\noindent Finally, let $G'$ be the graph obtained from $\widehat{G}$ after contracting every connected component $H$ of $\widehat{G}\setminus (C_{m}\cup C_{g})$ 
into a single vertex $v_{H}$ and setting as $\displaystyle w(v_{H})=\sum_{v\in V(H)}w(v)$. We define $I$ to be the set $V(G')\setminus (C_{m}\cup C_{g})$. 
Then~(\ref{clse3}) follows from construction of $I$. Claim 4 implies that for every vertex $v\in I$, $N_{G'}(v)\subseteq C_{m}$. 
This completes the proof of~(\ref{clse4}) and concludes the construction of the instance 
$(G',k)$ and the identification of the sets $C_{m}$, $C_{g}$, and $I$.\\

\noindent 
It remains to prove that the instances $(G,k)$ and $(G',k)$ are equivalent. 
As it has already been proved that the instances $(G,k)$ and $(\widehat{G},k)$ are equivalent
it is enough to prove that $(\widehat{G},k)$ and $(G',k)$ are equivalent.
Notice that if $(G',k)$ is a yes instance then $(\widehat{G},k)$ is also a yes instance; for every vertex $v$ in the tree deletion set of weight at most $k$ of $G'$
we consider the vertex $v$ in the tree deletion set of $\widehat{G}$ whenever $v\in C_{m}\cup C_{g}$ 
and the vertices of the connected component that was contracted to $v$ whenever $v\in I$.  
In order to prove that if $(\widehat{G},k)$ is a yes instance then $(G',k)$ is also a yes instance we start with the 
following. \\

\noindent {\em Claim 5.} If there exists a tree deletion set $S$ of $\widehat{G}$ of weight at most $k$ then there exists a tree deletion set $\widehat{S}$ of $\widehat{G}$ 
of weight at most $k$ such that for every $H\in \cc(\widehat{G}\setminus (C_{m}\cup C_{g}))$, either $V(H)\subseteq \widehat{S}$ or $V(H)\cap \widehat{S}=\emptyset$.\\

\noindent {\em Proof of Claim 5.} Recall that for every $H\in \cc(\widehat{G}\setminus (C_{m}\cup C_{g}))$ 
it holds that $N_{\widehat{G}}(V(H))$ is a double clique (Claim 4).
Therefore, either $N_{\widehat{G}}(V(H))\subseteq S$ or there exists a unique vertex of $N_{\widehat{G}}(V(H))$ that does not belong to $S$.
Notice that in the case where $N_{\widehat{G}}(V(H))\subseteq S$, as $N_{\widehat{G}}(V(H))$ is a separator of $\widehat{G}$, 
it trivially follows that either $V(H)\subseteq S$ or $S\subseteq\widehat{G}\setminus V(H)$ and the claim holds. Let us now assume that there exists a unique vertex $w$ of 
$N_{\widehat{G}}(V(H))$ that does not belong to $S$ and that $V(H)\cap S\neq \emptyset$. As, from Claim 1, every vertex of $N_{\widehat{G}}(V(H))$ has exactly
one neighbor in $H$ it follows that the graph $\widehat{G}[V(H)\cup \{w\}]$ does not contain a cycle.
Moreover, $w$ is a cut vertex of $\widehat{G}\setminus S$ and therefore the graph $(\widehat{G}\setminus S)\cup \widehat{G}[V(H)\cup \{w\}]$ is a tree.
Thus in the case where $V(H)\cap S\neq \emptyset$ we can remove the vertices of $V(H)$ from $S$ 
without introducing any cycles to the graph $(\widehat{G}\setminus S)\cup \widehat{G}[V(H)\cup \{w\}]$.
Therefore $S\setminus V(H)\subseteq S$ is also a tree deletion set of $\widehat{G}$ and this concludes the proof of the claim.
\hfill $\diamond$\\

Let now $\widehat{S}$ be a tree deletion set of $\widehat{G}$ of weight at most $k$. From Claim 5 we may assume that for every connected component $H$ of 
$\widehat{G}\setminus (C_{m}\cup C_{g})$ either $V(H)\subseteq \widehat{S}$ of $V(H)\cap \widehat{S}=\emptyset$.
Then it is straightforward to see that the vertex set $S$ consisting of the vertices $(C_{m}\cup C_{g})\cap\widehat{S}$ and the vertices of $I$ that correspond to the
connected components of $\widehat{G}\setminus (C_{m}\cup C_{g})$ whose vertices belong to $\widehat{S}$ is a tree deletion set of $G'$ of weight equal to the weight 
of $\widehat{S}$.
\end{proof}

\subsection{Results on Linear Equations}\label{subsct:lnreqtns}

\begin{lemma}\label{mtrx1lem}
For every matrix $M\in\mathbb{R}^{m\times n}$ and positive integer $k$, there exists a submatrix $M'\in \mathbb{R}^{m'\times n}$ of $M$, 
where $m'\leq n(k+1)$, such that for every $x\in \mathbb{R}^{n}$ with $\hd(M'\cdot x^{T},\mathbf{0})\leq k$, $\hd(M\cdot x^{T},\mathbf{0})=\hd(M'\cdot x^{T},\mathbf{0})$. 
Furthermore, the matrix $M'$ can be computed in time $O(m\cdot n^{\omega-1}k)$, where $\omega$ is the matrix multiplication exponent ($\omega<2.373$~\cite{Williams12}).
\end{lemma}

\begin{proof}
In order to identify $M'$ we identify $j_{0}+1\leq k+1$ (non-empty) submatrices of $M,B_{0},B_{1},\dots,B_{j_{0}}$ that have at most $n$ rows each 
 in the following way:
First, let $B_{0}$ be a minimal submatrix of $M$ whose rows span all the rows of $M$, that is, let $B_{0}$ be a base of the 
vector space generated by the rows of $M$, and 
let also $M_{0}$ be the submatrix obtained from $M$ after removing the rows of $B_{0}$.
We identify the rest of the matrices inductively as follows: For every $i\in [k]$, if $M_{i-1}$ is not the empty matrix we let $B_{i}$ 
be a minimal submatrix of $M_{i-1}$ whose rows span all the rows of $M_{i-1}$ and finally we let $M_{i}$ be the matrix occurring 
from $M_{i-1}$ after removing the rows of $B_{i}$.

We now define the submatrix $M'$ of $M$. Let $j_{0}\leq k$ be the greatest integer for which $M_{j_{0}-1}$ is not the empty matrix.
Let $M'$ be the matrix consisting of the union of the rows of the (non-empty) matrices $B_{0}$ and $B_{i}$, $i\in [j_{0}]$. 
As the rank of the matrices $M$, $M_{i}$, $i\in [j_{0}]$, is upper bounded by 
$n$, the matrices $B_{0}$, $B_{i}$, $i\in [j_{0}]$, have at most $n$ rows each, and therefore $M'$ has at most $n(j_{0}+1)\leq n(k+1)$ rows. 
Observe that if $j_{0}<k$ then the union of the rows of the non-empty matrices $B_{0}$, $B_{i}$, $i\in [j_{0}]$,
contains all the rows of $M$ and thus we may assume that $M'=M$ and the lemma trivially holds.
Hence, it remains to prove the lemma for the case where $j_{0}=k$, and therefore $M'$ consists of the union of the matrices $B_{0},B_{i}$, $i\in [k]$.
 As it always holds that 
 $$\hd(M\cdot x^{T},\mathbf{0})\geq\hd(M'\cdot x^{T},\mathbf{0})$$ 
 it is enough to prove that for every $x\in \mathbb{R}^{n}$ for which 
 $\hd(M'\cdot x^{T},\mathbf{0})\leq k$, $\hd(M\cdot x^{T},\mathbf{0})\leq\hd(M'\cdot x^{T},\mathbf{0})$. Thus, it is enough to prove that for every row $r$ 
 of the matrix $M''$ obtained from $M$ after removing the rows of $M'$, it holds that $\hd(r\cdot x^{T},\mathbf{0})=0$. 
 Towards this goal let $x\in \mathbb{R}^{n}$ be a vector such that $\hd(M'\cdot x^{T},\mathbf{0})\leq k$. 
 From the Pigeonhole Principle there exists an $i_{0}$ such that $\hd(B_{i_{0}}\cdot x^{T},\mathbf{0})=0$,
 that is, if $r_{1},r_{2},\dots,r_{|B_{i_{0}}|}$ are the rows of $B_{i_{0}}$ then $r_{j}\cdot x^{T}=0$, for every $j\in [|B_{i_{0}}|]$.
Recall however that the row $r$ of $M''$ is spanned by the rows $r_{1},r_{2},\dots,r_{|B_{i_{0}}|}$ of $B_{i_{0}}$. 
Therefore, there exist $\lambda_{j}\in \mathbb{R}$, $j\in [|B_{i_{0}}|]$, such that $\displaystyle r=\sum_{j\in [|B_{i_{0}}|]}\lambda_{j}r_{j}$.
It follows that 
$\displaystyle r\cdot x^{T}=\sum_{j\in [|B_{i_{0}}|]}\lambda_{j} (r_{j}\cdot x^{T})=0$ 
and therefore $\hd(r\cdot x^{T},\mathbf{0})=0$.
This implies that
$\hd(M\cdot x^{T},\mathbf{0})\leq\hd(M'\cdot x^{T},\mathbf{0}).$
Finally, for a rectangular matrix of size $d\times r$, $d\geq r$, Bodlaender et al.~\cite{BodlaenderCKN13} give an algorithm that computes a minimum weight column basis in
time $\Oh(dr^{\omega-1})$. By running this algorithm $k+1$ times we can find the matrix $M'$ in time $\Oh(mn^{\omega-1}k)$ and this completes the proof
of the lemma. 
\end{proof}

\begin{lemma}\label{mtrx2lem}
There exists an algorithm that given a set ${\cal S}$ of linear equations on $n$ variables and an integer $k$
outputs a set ${\cal S}'\subseteq {\cal S}$ of at most $(n+1)(k+1)$ linear equations such that any assignment of the variables that violates at most $k$ linear
equations of ${\cal S'}$ satisfies all the linear equations of ${\cal S}\setminus {\cal S'}$. Moreover, the running time of the algorithm is $\Oh(|{\cal S}| n^{\omega-1}k)$. 
\end{lemma}

\begin{proof}
Let $x_{1},x_{2},\dots,x_{n}$ denote the $n$ variables and $\alpha_{ij}$ denote the coefficient of $x_{j}$ in the $i$-th linear equation of $S$, $i\in [|{\cal S}|]$, $j\in [n]$. 
Let also $\alpha_{i(n+1)}$ denote the constant term of the $i$-th linear equation of ${\cal S}$. In other words, the $i$-th equation of ${\cal S}$ is denoted as
$\alpha_{i1}x_{1}+\alpha_{i2}x_{2}+\dots+\alpha_{in}x_{n}+\alpha_{i(n+1)}=0$.
Finally, let $M$ be the matrix where the $j$-element of the $i$-th row is 
$\alpha_{ij}$, $i\in [|{\cal S}|]$, $j\in [n+1]$. From Lemma~\ref{mtrx1lem}, it follows that
for every positive integer $k$ there exists a submatrix $M'$ of $M$ with at most $(n+1)(k+1)$ rows and $n+1$ columns such that for every 
$x\in \mathbb{R}^{n+1}$ for which $\hd(M'\cdot x^{T},\mathbf{0})\leq k$, $\hd(M\cdot x^{T},\mathbf{0})=\hd(M'\cdot x^{T},\mathbf{0})$ and
$M'$ can be computed in time $\Oh(|{\cal S}|n^{\omega -1}k)$. 
Let ${\cal S'}$ be the set of linear equations that correspond to the rows of $M'$. Let then $x_{i}=\beta_{i}$, 
$\beta_{i}\in \mathbb{R}$, $i\in [n]$, be an assignment that does not satisfy at most $k$ of the equations of ${\cal S'}$.
This implies that $\hd(M'\cdot z,\mathbf{0})\leq k$, where $z=(\beta_{1},\beta_{2},\dots,\beta_{n},1)^{T}$. Again, from Lemma~\ref{mtrx1lem}, 
we get that $\hd(M\cdot z,\mathbf{0})=\hd(M'\cdot z,\mathbf{0})$. Thus, the above assignment satisfies all the
linear equations of ${\cal S}\setminus {\cal S'}$.
\end{proof}

\subsection{The Main Theorem}

In this subsection by combining the structural decomposition of Subsection~\ref{subsct:strcdecomp} 
and Lemma~\ref{mtrx2lem} from Subsection~\ref{subsct:lnreqtns} we obtain a kernel for {\sc wTDS} of size $\Oh(k^{4})$.

\begin{theorem} 
{\sc wTDS} admits a kernel of size $\Oh(k^{4})$ and  $\Oh(k^{4}\log k)$ bits.
\end{theorem}
\begin{proof}
Let $(G,k)$ be an instance of {\sc wTDS}. Without loss of generality we may assume that it is semi-reduced and that, 
from Lemma~\ref{mainlem}, $V(G)$ can be partitioned into 
three sets $C_{m}$, $C_{g}$, and $I$ satisfying the conditions of Lemma~\ref{mainlem}. Note here that as $G$ is semi-reduced, $G$ is connected and therefore
every vertex of $I$ has at least one neighbor in $C_{m}$.
We construct an instance $(G',k)$ of {\sc wTDS} in the following way.
Let $I=\{v_{i}\mid i\in[|I|]\}$ and $C_{m}=\{u_{j}\mid j\in [|C_{m}|]\}$. We assign a variable $x_{j}$ to $u_{j}$, $j\in [|C_{m}|]$, and a linear equation $l_{i}$
to $v_{i}$, $i\in [|I|]$, where $l_{i}$ is the equation $\displaystyle\sum_{j\in [|C_{m}|]}\alpha_{ij}x_{j}-1=0$ and $\alpha_{ij}=1$ if $u_{j}\in N_{G}(v_{i})$ 
and 0 otherwise. Let ${\cal L}=\{l_{i}\mid i\in [|I|]\}$ and ${\cal L}'$ be the subset of ${\cal L}$ obtained from Lemma~\ref{mtrx2lem}. 
Let also $I' = \{v_{p}\in I\mid l_{p}\in {\cal L}'\}$ and $G'=G[C_{m}\cup C_{g}\cup I']$. We now prove that $(G',k)$ is equivalent to $(G,k)$. 

We first prove that if $(G,k)$ is a yes instance then so is $(G',k)$.
Let $S$ be a tree deletion set of $G$ of weight at most $k$. Then $G\setminus S$ is a tree and as for every vertex $v\in I\setminus S$, 
$G[N_{G}(v)]$ is a double clique then $v$ has degree exactly 1 in $G\setminus S$. Therefore, the graph obtained from $G\setminus S$ 
after removing $(I\setminus I')$ is still a tree. This implies that $S\setminus (I\setminus I')$ is a tree deletion
set of $G'$ of weight at most $k$ and $(G',k)$ is a yes instance.

Let now $(G',k)$ be a yes instance and $S$ be a tree deletion set of $G'$ of weight at most $k$. 
We claim that there exist at most $k$ vertices 
in $I'$ whose neighborhood lies entirely in $S$. Indeed, assume to the contrary that there exist at least $k+1$ vertices of $I'$ whose neighborhood lies
entirely in $S$. Let $J$ be the set of those vertices. Notice that for every vertex $v\in I'$, if $N_{G'}(v)\subseteq S$, then either $v\in S$ or 
$I' \setminus \{v\}\subseteq S$.
Notice that if $J\subseteq S$, then $S$ has weight at least $k+1$, a contradiction. Therefore, there exists a vertex $w\in J$ that is not contained in $S$.
Then $I'\setminus \{w\}\subseteq S$.  Moreover, recall that $w$ has at least one neighbor $z$ in $C_{m}$ and from the hypothesis $z$ is contained in $S$. 
Therefore $(I'\setminus \{w\})\cup \{z\}\subseteq S$. As $|I'|\geq |J|=k+1$, it follows that $|I'\setminus \{w\}| \geq k$. Furthermore, recall that 
$C_{m}\cap I'=\emptyset$. Thus, $|S|\geq k+1$, a contradiction to the fact that $S$ has weight at most $k$.
Therefore, there exist at most $k$ vertices of $I'$ whose neighborhood is contained entirely in $S$.
For every $j\in [|C_{m}|]$, let $x_{j}=\beta_{j}$, where $\beta_{j}=0$ if $u_{j}\in S$ and 1 otherwise.
Then there exist at most $k$ linear equations in ${\cal L}'$ which are not satisfied by the above assignment. 
However, from the choice of ${\cal L}'$ all the linear equations in ${\cal L}\setminus {\cal L}'$ are satisfied and therefore, every vertex in $I\setminus I'$ has exactly one 
neighbor in $G\setminus S$. Thus $G\setminus S$ is a tree and hence, $S$ is a tree deletion set of $G$ as well.

Notice that $V(G')=C_{m}\cup C_{g} \cup I'$, where $|I'|\leq 8k^{3}+10k^{2}+3k+1$ (Lemma~\ref{mtrx2lem}) and therefore $|V(G')|=\Oh(k^{4})$.
It is also easy to see that $|E(G')|=\Oh(k^{4})$. Indeed, notice first that as the set $I'$ is an independent set there are no edges between its vertices. Moreover,
from Lemma~\ref{mainlem} there are no edges between the vertices of the set $I'$ and the set $C_{g}$. Observe that, from the construction of $I$ and subsequently
of $I'$, Lemma~\ref{mainlem} implies that every vertex of $I'$ has at most $2k+1$ neighbors in $C_{m}$. As $|I'| \leq 8k^{3}+10k^{2}+3k+1$ there exist $\Oh(k^{4})$
edges between the vertices of $I'$ and the vertices of $C_{m}$. Notice that from~(\ref{clse2}) of Lemma~\ref{mainlem}, $C_{g}$ induces a forest and thus there 
exist at most $\Oh(k^{4})$ edges between its vertices. Moreover, from~(\ref{clse1}) of Lemma~\ref{mainlem}, again there exist $\Oh(k^{4})$ edges between the 
vertices of $C_{m}$. It remains to show that there exist $\Oh(k^{4})$ edges with one endpoint in $C_{m}$ and one endpoint in $C_{g}$. Recall first that 
every $x\in C_{m}$ has at most one neighbor in every  connected component of $G[C_{g}]$ and that every connected component has at most 2 neighbors in
$Q$. Therefore, there exist at most $2k+2$ edges between every connected component of $G[C_{g}]$ and $C_{m}$. 
Moreover, from Eq.~\eqref{grbgdfn},\eqref{eqngergkmlerg},~\eqref{aefalwefnlkflwa}, and~\eqref{aergwalag}, and Claim 3 we obtain that $C_{g}$ contains 
$\Oh(k^{3})$ connected components. Therefore, there exist $\Oh(k^{4})$ edges with one endpoint in $C_{m}$ and one endpoint in $C_{g}$.
Thus, {\sc wTDS} has a kernel of $\Oh(k^{4})$ vertices and edges. 
Finally, from Reduction Rule~\ref{rrule6}, the weight of every vertex is upper bounded by $k+1$ and thus, it can be encoded using $\log (k+1)$ bits resulting to a
kernel of {\sc  wTDS} with $\Oh(k^{4}\log k)$ bits.
\end{proof}
\newpage

\bibliography{tdsbibl,tds-final}

\end{document}